\documentclass[twocolumn,showpacs,aps,prb,letterpaper,nofootinbib,raggedbottom,nobalancelastpage]{revtex4}
\usepackage{graphicx,bm}
\usepackage{amsmath}
\usepackage{color}
\begin{document}
\title{Fast generation of entangled photon pairs from a single quantum dot embedded in a photonic crystal cavity}
\date{\today}
\author{P. K. Pathak and S. Hughes}
\address{Department of Physics,
Queen's University, Kingston, ON K7L 3N6, Canada}
\begin{abstract}
We present a scheme for the fast generation of entangled photons from a single quantum dot coupled
to a planar photonic crystal that support two orthogonally polarized cavity modes. We discuss ``within generation'' and ``across generation'' of entangled photons when both biexciton to exciton, and exciton to ground state transitions, are coupled through cavity modes. In the across generation, the photon entanglement is restored through a time delay between the photons. The two photon concurrence, which is
 a measure of entanglement, is greater than 0.7 and 0.8 using experimentally achievable
parameters in across generation and within generation, respectively. We also show that the entanglement can be distilled in both cases using a simple spectral filter.

\end{abstract}
\pacs{03.65.Ud, 03.67.Mn, 42.50.Dv}
\maketitle

\section{Introduction}
Entangled photons are an essential resource for various quantum information processing protocols~\cite{qcomp,qcomp2}, such as quantum cryptography~\cite{crypto} and
quantum teleportation~\cite{tele}. The entangled photons employed in most experiments to date have been generated using parametric down conversion~\cite{mandelwolf,kwait}. However, recent developments of scalable quantum systems~\cite{scalable} will require a scalable ``on demand'' source of entangled photons.

With regard to suitable material systems for on demand photon sources,
there has been considerable progress for developing entangled photon sources using single quantum dots (QDs)~\cite{akopian,mfield,efield,annealing,dotselection,timegate}. In
semiconductor QDs, entangled photons are generated in a biexciton-exciton cascade decay. However, the entanglement between the generated photons is limited by inherent cylindrical asymmetries and various dephasing processes\cite{dephasing,dephasing2,new_tejedor}. The cylindrical asymmetries produce fine structure splitting (FSS) in the exciton states~\cite{anisotropy}; as a result, the emitted $x-$polarized and $y-$polarized photon pairs become distinguishable in frequencies, and the entanglement between the photons is largely destroyed. Several methods have been employed to minimize the detrimental
effects of FSS on generated photons, for example, by spectrally filtering  the indistinguishable photon pairs~\cite{akopian}, by applying external fields to suppress the FSS~\cite{mfield,efield}, by thermal annealing the QDs~\cite{annealing}, by selecting QDs with smaller FSS~\cite{dotselection}, and by using temporal gates~\cite{timegate}. In all of these approaches, the photons of different polarizations, generated within the same generations, are forced to match in their frequencies.

\begin{figure}[b!]
\centering
\includegraphics[width=3.2in]{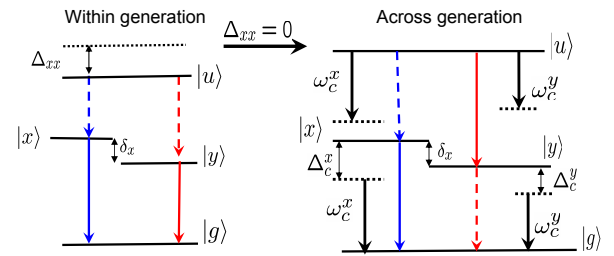}
\caption{(Color online) Schematic of
the resulting energy level diagram for cavity-QED assisted generation of entangled photons in the biexciton-exciton
cascade decay after manipulating the binding energy of biexciton ($\Delta_{xx} \rightarrow 0$). The biexciton state $|u\rangle$ decays to the ground state $|g\rangle$ via intermediate exciton state $|x\rangle$ or $|y\rangle$, creating an $x-$polarized or $y-$polarized photons in the cavity modes. The $x-$polarized and $y-$polarized cavity modes are coupled with the $|u\rangle\rightarrow|x\rangle$, $|x\rangle\rightarrow|g\rangle$ and $|u\rangle\rightarrow|y\rangle$, $|y\rangle\rightarrow|g\rangle$ transitions, respectively.}
\label{fig1}
\end{figure}

An interesting alternate approach, insensitive to FSS, has been proposed recently, which suppressing the binding energy of the biexciton~\cite{reimer,avron,tejedor}. For a zero binding energy of the biexciton, photons of different polarizations match in energy in ``across generations'' (see Fig.~\ref{fig1}). Because of the different ordering in the emission for $x-$polarized and $y-$polarized photon pairs, the photons are distinguishable temporarily and remain unentangled, but the entanglement can be
restored using a time delay between photons of different generations.

The effects of dephasing in the generated entangled state of photons can be minimized significantly by
enhancing the emission rates of photons through the Purcell effect in a system comprised of a QD coupled with a microcavity.
Several experiments have also demonstrated
 single QD strong coupling to semiconductor cavities~\cite{photocavity1,photocavity2,Peter2005}.
Recently, Johne {\em et al.}\,\cite{johne} proposed a cavity-QED scheme for generating entangled photons in the strong coupling regime. In their scheme, excitons are strongly coupled with cavity modes and form degenerate polariton states~\cite{polariton}. A formal theory of this scheme, including
  exciton and biexciton broadenings, has been reported by us~\cite{ours}. However,
  one drawback of the proposed scheme is that because of the large binding energy, the biexciton remains uncoupled with cavity modes and thus the first generation of photons has a
long life time.
 In this paper, we propose a scheme for the fast generation of entangled photons from a single QD, by manipulating the binding energy of the biexciton such that both biexciton to excitons and excitons to ground state are coupled with two cavity modes of orthogonal polarization. Experimentally,
manipulation of the binding energies of the biexciton has been reported by applying lateral electric field~\cite{reimer} and by thermal annealing~\cite{annealing}. In this proposed
fast generation schemes introduced below, we discuss both ``across generation'' and ``within generation'' of entangled photons.
%using the same material, thus
%merging the ideas of cavity-QED and biexciton binging energy suppression to enhance the
%efficiency of the entangled photon pair generation.

This paper is organized as follows. In Sec.~II, we present a formal theory of a single QD coupled to a
 planar photonic cavity cavity. The cavity-assisted {\em across generation} of entangled photons is discussed in Sec.~III. In Sec.~IV, the cavity-assisted {\em within generation} of entangled photons  is studies. In section~V, we present our conclusions. In the appendix, we show a derivation for the
 dressed states of the biexciton.

\section{Theory}
We consider a QD embedded in a photonic crystal (PC) cavity having two orthogonal polarization modes of frequencies $\omega_c^x$ and $\omega_c^y$, which can be realized and tuned experimentally using
electron-beam lithography and, for example, AFM oxidization techniques \cite{HennessyAPL06}. The exciton states, $|x\rangle$ and $|y\rangle$, have FSS $\delta_{\rm x}$. The cavity modes are coupled with the  biexciton to exciton and exciton to ground-state transitions,
by manipulating the biexciton binding energy~\cite{annealing,reimer}.
The schematic arrangement of the system is shown in Fig.~1.

The Hamiltonian for the system of QD coupled with two-modes in PC-cavity, in the interaction picture, can be  written as
%\begin{eqnarray}
\begin{widetext}
\begin{eqnarray}
\frac{H_{I}(t)}{\hbar} &  = &  g^x_1|x\rangle\langle g|\hat a_c^x e^{i\Delta_c^x t}
+g^x_2|u\rangle\langle x|\hat a_c^x e^{i(\omega_{ux}-\omega_c^x) t}
+g^y_1|y\rangle\langle g|\hat a_c^y e^{i\Delta_c^y t}+g^y_2|u\rangle\langle y|\hat a_c^y e^{i(\omega_{uy}-\omega_c^x) t}
 \nonumber \\
& \hbox{} & + \sum_{m\ne c}\Omega_{xm}\hat a_c^{x\dag}\hat a^x_m e^{i(\omega_c^x-\omega_m)t}
 +\sum_{m\ne c}\Omega_{ym}\hat a_c^{y\dag}\hat a^y_m e^{i(\omega_c^y-\omega_m)t}
+ H.c., \,
%\end{eqnarray}
\end{eqnarray}
\end{widetext}
%plus a Hermitian conjugate term,
where $\omega_{ux}=\omega_u-\omega_x$, $\omega_{uy}=\omega_u-\omega_y$, $\Delta_c^x=\omega_x-\omega^x_c$, $\Delta_c^y=\omega_y-\omega_c^y,$
and $\hat a^i_j$ are the field operators with  $\hat a_c^x$ and $\hat a_c^y$ the cavity mode operators.
Here, $\Omega_{xm}$, and $\Omega_{ym}$ represent the couplings to the environment from
the cavity mode; $g^i_j$ are the coupling strength between the
exciton/biexciton and cavity mode; $\omega_m^i$ are the frequencies of the $i-$polarized photons emitted from the cavity mode, and $\omega_u$, $\omega_x$, $\omega_y$
are the the frequency of the biexciton and excitons, respectively.
We consider a system that is optically pumped in such a way as to have
an initially-excited biexciton, with no photons inside the cavity, thus
the state of the system at any time $t$ can be written as follows:
\begin{widetext}
%\begin{equation}
%\begin{minipage}[b]{\linewidth}
\begin{eqnarray}
|\psi(t)\rangle&  = &c_1(t)|u,0,0\rangle+c^x_2(t)|x,1,0\rangle+c^y_2(t)|y,0,1\rangle
%\nonumber\\
+c^x_3(t)|g,2,0\rangle+c^y_3(t)|g,0,2\rangle \nonumber \\
& \hbox{} &
+\sum_mc^x_{4m}(t)|x,0,0\rangle|1_m\rangle_x|0\rangle_y
+\sum_mc^y_{4m}(t)|y,0,0\rangle|0\rangle_x|1_m\rangle_y
+\sum_mc^x_{5m}(t)|g,1,0\rangle|1_m\rangle_x|0\rangle_y
 \nonumber \\
& \hbox{} &
+\sum_mc^y_{5m}(t)|g,0,1\rangle|0\rangle_x|1_m\rangle_y
+\sum_{mn}c^x_{mn}(t)|g,0,0\rangle|1_m,1_n\rangle_x|0\rangle_y
+\sum_{mn}c^y_{mn}(t)|g,0,0\rangle|0\rangle_x|1_m,1_n\rangle_y. \ \ \
\end{eqnarray}
\end{widetext}
The different terms in the state vector $|\psi\rangle$ represent, respectively:
 the dot is in the biexciton state with zero photons in the cavity; the dot is in the exciton state with one photon in the $x$-polarized cavity mode; the dot is in the exciton state with one photon in the $y$-polarized cavity mode; the dot is in ground state with two photon in $x$-polarized cavity mode; the dot is in the ground state with two photons in $y$-polarized cavity modes; and the additional possible terms due to leakage of photons from the cavity modes to the reservoirs; the suffixs to the reservoir kets represent their polarization.

%It is important here to consider two photons in cavity reservoirs distinguishable, i.e. $m\neq n$, in order to facilitate linear interaction between the cavity mode and
%the reservoir; this means that we neglect two-photon nonlinear interactions~\cite{cdegenerate},
% such as $\hat a_c^2 (\hat a_m^\dagger)^2$, which is well justified for
% our system of interest.

By using the Schr\"odinger equation,
 applying the  Weisskopf-Wigner approximation \cite{wwa,raymer,yao}, and introducing biexciton and exciton broadenings,
 we derive the following equations of motion
 for the probability  amplitudes:
 \begin{eqnarray}
%\label{cn1}
\dot{c}_1(t)&=&-ig^x_2c^x_{2}(t)e^{i(\omega_{ux}-\omega_c^x)t}-ig^y_2c^y_{2}(t)e^{i(\omega_{uy}-\omega_c^y)t}\nonumber\\&&-\gamma_1c_1(t),\\
\label{cn1}
\dot{c^\alpha_2}(t)&=&-ig^\alpha_2c_{1}(t)e^{-i(\omega_{u\alpha}-\omega_c^\alpha)t}-ig^\alpha_1\sqrt{2}\,c^\alpha_{3}(t)
e^{i\Delta_c^\alpha t}\nonumber\\&&-\kappa c^\alpha_{2}(t)-\gamma_2c^\alpha_{2}(t),\\
\label{cn2}
%\dot{c^y}_{2}(t)&=&-ig^y_2c_{1}(t)e^{-i(\omega_{uy}-\omega_c^y)t}-ig^y_1\sqrt{2}c^y_{3}(t)e^{i\Delta_c^yt}\nonumber\\&&-\kappa %c^y_{2}(t)-\gamma_2c^y_{2}(t),\\
\dot{c^\alpha_3}(t)&=&-ig^\alpha_1\sqrt{2}c^\alpha_{2}(t)e^{-i\Delta_c^\alpha t}-2\kappa c^\alpha_{3}(t),\\
%\dot{c^y}_{3}(t)&=&-ig^y_1\sqrt{2}c^y_{2}(t)e^{-i\Delta_c^yt}-2\kappa c^x_{3}(t),\\
\label{cn3}
\dot{c}^\alpha_{4m} (t)&=&-ig^\alpha_1c^\alpha_{5m}e^{i\Delta_c^\alpha t}-i\Omega_{\alpha m}^{*} c^\alpha_{2}e^{-i(\omega_c^\alpha-\omega_m)t}\nonumber\\&&-\gamma_2c^\alpha_{4m}(t),\\
\label{cn4}
%\dot{c^y}_{4m}(t)&=&-ig^y_1c^y_{5m}e^{i\Delta_c^yt}-i\Omega_{ym}^{*}c^x_{2}e^{-i(\omega_c^y-\omega_m)t}\nonumber\\&&-\gamma_2c^y_{4m}(t),\\
\dot{c}^\alpha_{5m}(t)&=&-ig^\alpha_1c^\alpha_{4m}e^{-i\Delta_c^\alpha t}-i\Omega_{\alpha m}^*\sqrt{2}\,c^\alpha_{3}(t)
e^{-i(\omega_c^\alpha-\omega_m)t}\nonumber\\&&-\kappa c^\alpha_{5m}(t), \nonumber \\
%\dot{c^y}_{5m}(t)&=&-ig^y_1c^y_{4m}e^{-i\Delta_c^yt}-i\Omega_{ym}^*\sqrt{2}c^y_{3}(t)e^{-i(\omega_c^y-\omega_m)t}\nonumber\\&&-\kappa c^y_{5m}(t)
\label{cn5}
\dot{c}^\alpha_{mn}(t)&=&-i\Omega_{\alpha n}^*c^\alpha_{5m}e^{-i(\omega_c^\alpha-\omega_n)t},
\label{cn6}
\end{eqnarray}
where $\alpha=x$ or $y$, $\kappa=\pi|\Omega_{xm}|^2=\pi|\Omega_{ym}|^2$ is the half width
of the cavity modes (assuming uniform and equal coupling for $x$ and $y$),
  %(we assume the same leakage for both cavity modes) is the half width of the cavity mode and
and  $\gamma_1$, $\gamma_2$ are
the half widths of the  biexciton and exciton levels, respectively.
We note that  $\gamma_1$ and $\gamma_2$ can include
both  radiative and nonradiative broadening, and for QDs, $\gamma_1\approx2\gamma_2$.
We next solve Eqs.(\ref{cn1})-(\ref{cn5}) to obtain $c^x_{mn}$ and $c^y_{mn}$, using the Laplace transform method.
The probability amplitudes for emission of a photon pair, in the long time limit, are
\begin{eqnarray}
\label{fcmn}
c^x_{mn}(\infty) &  =  &\frac{g^x_1\Omega^*_{xn}(\omega_m+3\omega_n-2\omega_x-2\omega_c^x+2i\kappa+2i\gamma_2)}
{(\omega_n-\omega_x+i\gamma_2)(\omega_n-\omega_c^x+i\kappa)-(g^x_1)^2} \nonumber \\
  &  & \times \frac{g^x_2\Omega^*_{xm}F_y(\omega_m,\omega_n)}{D(\omega_m,\omega_n)},\\
\label{fcpmn}
c^y_{mn}(\infty) & = & \frac{g^y_1\Omega^*_{yn}(\omega_m+3\omega_n-2\omega_y-2\omega_c^y+2i\kappa+2i\gamma_2)}
{(\omega_n-\omega_y+i\gamma_2)(\omega_n-\omega_c^y+i\kappa)-(g^y_1)^2} \nonumber\\
&  & \times \frac{g^y_2\Omega^*_{ym}F_x(\omega_m,\omega_n)}{D(\omega_m,\omega_n)},
\end{eqnarray}
where
\begin{eqnarray}
\label{gpm}
F_\alpha(\omega_m,\omega_n)&=&2(g^\alpha_1)^2-(\omega_m+\omega_n-\omega_\alpha-\omega_c^\alpha
+i\kappa+i\gamma_2)\nonumber\\&&(\omega_m+\omega_n-2\omega_c^\alpha+2i\kappa), \\
D(\omega_m,\omega_n)&=&(\omega_m+\omega_n-\omega_u+i\gamma_1)F_xF_y\nonumber\\
&& + (g^x_2)^2F_y(\omega_m+\omega_n-2\omega_c^x+2i\kappa)\nonumber\\
&& + (g^y_2)^2F_x(\omega_m+\omega_n-2\omega_c^y+2i\kappa).
\end{eqnarray}
The optical spectrum of the generated $x$-polarized photon-pair is given by $S(\omega_m,\omega_n)=|c^x_{mn}(\infty)|^2$, and the spectrum for $y$-polarized photon pair is given by $S(\omega_m,\omega_n)=|c^y_{mn}(\infty)|^2$.
The spectral functions, $S(\omega_m,\omega_n)$,
represent the joint probability distribution, and thus the integration over the one frequency
variable gives the spectrum at the other frequency. For example, the spectrum of the first generation of photons emitted via cavity mode is given by $S(\omega_m)=\int_{-\infty}^{\infty} S(\omega_m,\omega_n)\,d\omega_n$, and the spectrum of second generation of photons is $S(\omega_n)=\int_{-\infty}^{\infty} S(\omega_m,\omega_n)\,d\omega_m$.

From the above discussion, the state of the {\em photon pair} emitted from both the cavity modes is given by
\begin{eqnarray}
|\psi\rangle=\sum_{m,n}c^{x}_{mn}(\infty)|1_m,1_n\rangle_x+\sum_{m,n}c^{y}_{mn}(\infty)|1_m,1_n\rangle_y ,
\label{state}
\end{eqnarray}
where in each term the ket represents the state of the cavity mode reservoirs, and the ket suffix  labels the polarization.
The coefficients $c^\alpha_{mn}(\infty)$ are given by the analytical
expressions described through Eqs.~(\ref{fcmn}) and (\ref{fcpmn}).

\section{Cavity-assisted ``across generation'' of entangled photons}
In the previous section, we have derived expressions for the final state of the photons generated in the biexciton-exciton cascade decay through leaky cavity modes. Depending on the coupling strength and detunings of the cavity modes from the transition frequencies in QD, the emitted $x-$polarized and $y-$polarized photons can match in energies {\em within} the same generations or through {\em across} generations. In this section we discuss the case when the photons match in energy in across generations.
The state of the emitted photon pair is given by
\begin{equation}
|\psi\rangle=\sum_{k,l}[c^x_{kl}(\infty)|1_k\rangle_x|1_l\rangle_x+c^y_{lk}(\infty)|1_l\rangle_y|1_k\rangle_y] ,
\label{ac}
\end{equation}
where the first and second ket in each term show the photon of the first generation and the second generation, respectively; the second term corresponding to the $y-$polarized photon pair has
the reverse ordering of indices compared to the first term.
%The coefficients c are given by Eqs.~(\ref{fcmn})-(\ref{fcpmn}).
 Although the photons of different polarizations in different generation could be degenerate in frequencies, they are distinguishable in order, namely in time. Thus, for generating entangled photons it is necessary to make photons temporally indistinguishable as well. For erasing the temporal information, photons of the first generation are delayed by time $t_0$. The normalized off-diagonal element of the density matrix of photons, in the polarization basis, is given by
\begin{equation}
\gamma=\frac{\int\int c^{x*}_{kl}(\infty)c^y_{lk}(\infty)W_{\rm opt}(\omega_k,\omega_l)\,d\omega_kd\omega_l}{\int\int |c^{x}_{kl}(\infty)|^2\,d\omega_kd\omega_l+|c^y_{lk}(\infty)|^2\,d\omega_kd\omega_l},
\label{gmf}
\end{equation}
where $W_{\rm opt}=\exp[-i(\omega_k-\omega_l)t_0]$ is an additional phase generated by the time delay. For $t_0=0$, i.e., no time delay is employed, $W_{\rm opt}=1$, and from Eq.~(\ref{gmf}) one gets $\gamma=0$. This shows that the phase $W_{\rm opt}$ is essential to erase the
temporal information of photon emission from the state $\psi$ (Eq.~\ref{ac}). For a certain value of delay $t_0$, the photons of the first generation and second generations can become indistinguishable and the value of $|\gamma|$ has a maximum. We note that the concurrence~\cite{wooters}, which is a quantitative measure of entanglement, for the generated state of photons $|\psi\rangle$ is equal to $2|\gamma|$; so $|\gamma|=0.5$ represents the maximum entanglement.

In order to better understand the results for cavity-assisted generation of entangled photons we first consider the case when
the QD is not coupled with the cavity modes. In that case, the photons are generated in the spontaneous emission through biexciton-exciton cascade decay~\cite{avron}, and the coefficients $c$ in Eq.(\ref{ac}) are given by
\begin{eqnarray}
c^x_{kl}=\frac{\sqrt{\gamma_1\gamma_2/2\pi^2}}{(\omega_k+\omega_l-\omega_u+i\gamma_1)(\omega_l-\omega_{x}+i\gamma_2)},\\
c^y_{lk}=\frac{\sqrt{\gamma_1\gamma_2/2\pi^2}}{(\omega_k+\omega_l-\omega_u+i\gamma_1)(\omega_k-\omega_{y}+i\gamma_2)}.
\end{eqnarray}
For a QD having zero biexciton binding energy, i.e., $\omega_u=\omega_x+\omega_y$, and with a time delay $t_0$, from Eq.~(\ref{gmf}), one gets
\begin{equation}
\gamma=\frac{2\gamma_2 e^{-2\gamma_2 t_0}}{\gamma_1}(1-e^{-\gamma_1 t_0}).
\label{result}
\end{equation}
From Eq.~(\ref{result}), we notice that $\gamma$ is maximized for $\gamma_2 t_0=\gamma_2\ln(1+\gamma_1/2\gamma_2)/\gamma_1$. Normally for a QD, $\gamma_1/\gamma_2=2$,  and the maximum value of $\gamma$ is 0.25; so after manipulating the decay such that $\gamma_1/\gamma_2\rightarrow0$, the maximum value of $\gamma=1/e$ is obtained. Similar values have also been reported by simulating correlations within the density matrix formalism~\cite{tejedor}.

\begin{figure}[t]
\centerline{\includegraphics[width=3in]{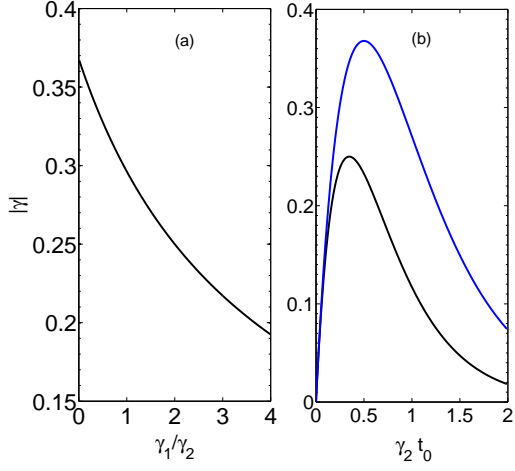}}
\caption{(color online)
(a) Optimum value of $|\gamma|$ corresponding to a time delay $\gamma_2 t_0=\gamma_2\ln(1+\gamma_1/2\gamma_2)/\gamma_1$. (b) The value of $|\gamma|$ for $\gamma_1/\gamma_2=2$ (black) and for $\gamma_1/\gamma_2\rightarrow0$ (blue).}
\label{fig2}
\end{figure}

It is important to note here, that the values of $|\gamma|$ using a time delay are quite different to
the values reported by Avron {\em et. al.}~\cite{avron}.
 The reason for this discrepancy, is that we have considered an experimentally feasible linear time delay,
 while Avron {\em et al.} considered a complex nonlinear time delay that
 is
%
% however Avron et. al. consider nonlinear time delay, which is almost
practically impossible to realize~\cite{comment}.
Consequently,
%Thus,
the maximum
%possible
value of concurrence in across generation of entangled photons through time reordering is 0.73, even after optimally manipulating the exciton/biexciton line widths. In Fig.~\ref{fig2}(a) we show the dependence of entanglement on the value of $\gamma_1/\gamma_2$. The dependence of the off-diagonal element of of photon density matrix on delay time is shown in Fig.~\ref{fig2}(b). For QDs, $\gamma_1$ and $\gamma_2$ have radiative and non-radiative parts, and
generally the nonradiative parts are
%much
larger than the radiative parts. Thus it is not possible to manipulate the values of $\gamma_1/\gamma_2$ significantly by changing the decay rates of the biexciton and excitons~\cite{dotsize}. However, in the
coupld QD  - PC cavity system, the radiative halfwidths of biexciton and excitons can be significantly larger than their nonradiative half widths, and by tuning the cavity mode frequencies and couplings
 parameters one can manipulate the ratio of the biexciton line width to the exciton line width and thus
 increase the degree of entanglement. Also, the required delay time for maximizing the entanglement can be achieved by creating path differences for photons of selected polarization and frequency.  For smaller values of $\gamma_2$,
one must generate a large optical path difference between photons to realize the appropriate time delays $t_0$, corresponding to $\gamma_2 t_0=\gamma_2\ln(1+\gamma_1/2\gamma_2)/\gamma_1$. However, for a QD coupled with a cavity,
the decay rates of the biexciton and exciton could be very large, thus the required delay time will be significantly small and can be achieved easily
%easily
in an appropriate optical delay scheme~\cite{avron}.

\begin{figure}[t!]
\centerline{\includegraphics[width=3in, height=3.2in]{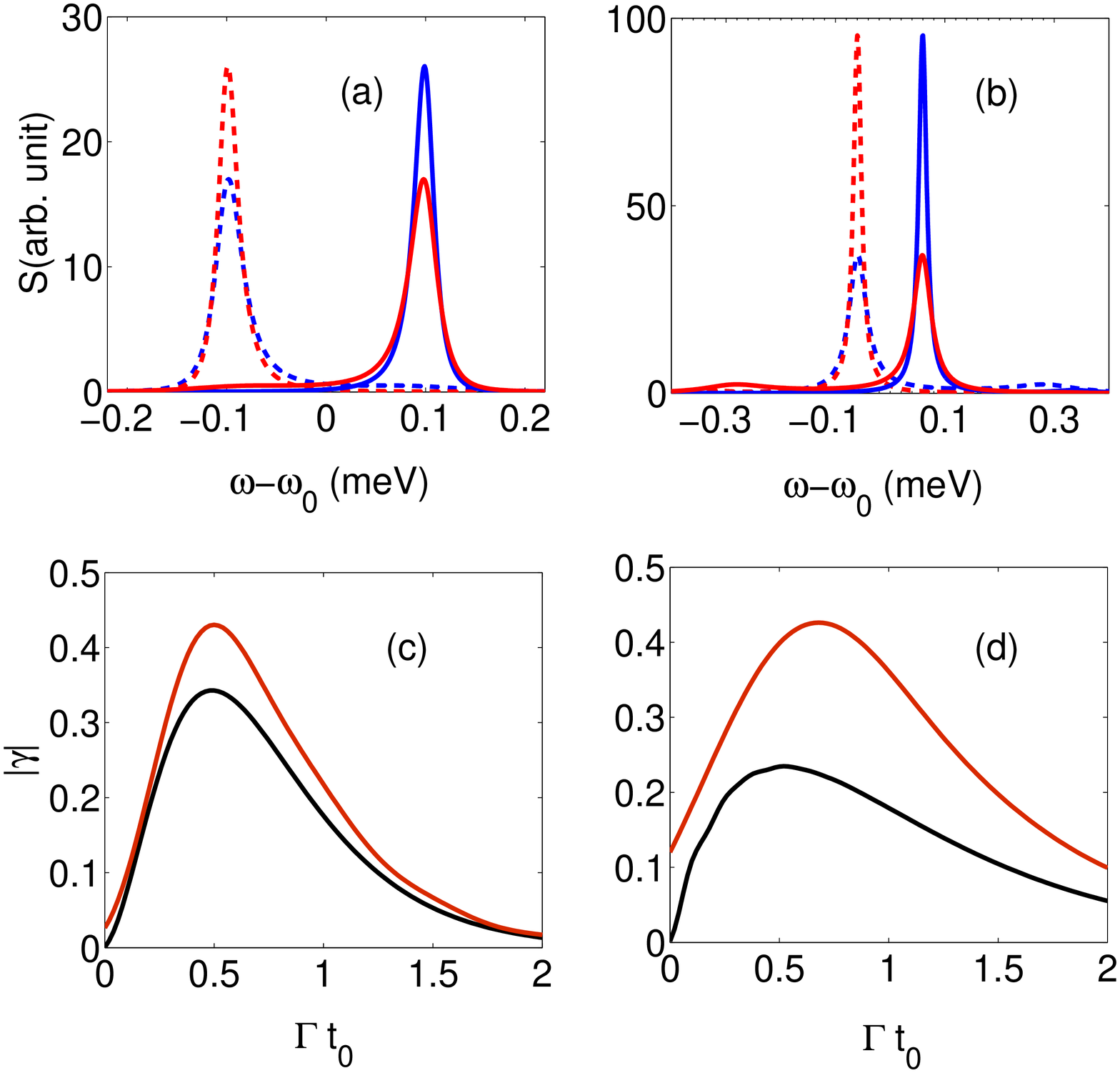}}
\vspace{-0.3cm}
\caption{(Color online)
The ``across generation'' of entangled photons when biexciton is also coupled with cavity modes after reducing binding energy $\Delta_{xx}=0$\,meV.
On the left, we consider weak coupling with the cavity modes degenerate with the
exciton modes, and on the right we consider strong coupling with the cavity modes
detuned with respect to the exciton modes.
The spectrum of the photons $S(\omega)$ for $\delta_x=0.2$\,meV, $\gamma_1=2\gamma_2=0.004\,$meV, $\kappa=0.05$\,meV, for (a) $g^x_1=g^x_2=g^y_1=g^y_2=0.02$\,meV, and $\Delta_c^x=\Delta_c^y=0$\,meV, and for (b) $g^x_1=g^x_2=g^y_1=g^y_2=0.1$\,meV, and $\Delta_c^x=-\Delta_c^y=-0.2$\,meV. The $x-$polarized photons are shown in blue and the $y-$polarized
are shown in red; also, the solid curves are for photons generated in the exciton decay and the dotted curves are for photons generated in the biexciton decay.
(c-d) The values of $|\gamma|$ corresponding to time delay $\Gamma t_0$, where $\Gamma=g^2\kappa/(\kappa^2+\Delta_c^{x2})$. The red (black) curves represent results for filtered (unfiltered) photons. For (c) the filter function corresponds to two spectral windows of width $w=0.05\,$\,meV, centered at $\omega_x$ and $\omega_y$, and for (d) the filter function corresponds to two spectral windows of width $w=0.03\,$\,meV, centered at $\omega_x^-$ and $\omega_y^+$.}
\vspace{-0.1cm}
\label{fig3}
\end{figure}

For {\em across generation} of entangled photons, we consider a QD coupled with a PC-cavity when the binding energy of biexciton is suppressed to zero. We plot values of $|\gamma|$ for typical values of cavity couplings and detunings in Fig.~\ref{fig3}. For the
weak coupling regime, the radiative decay rates of the exciton states via the cavity modes are given by $\Gamma^i_2=g^{i~2}_1\kappa/(\kappa^2+\Delta^{i~2}_{c})$, for $i=x,y$. The radiative decay rates for the biexciton state $|u\rangle$ into the exciton states $|x\rangle$ and $|y\rangle$ are given by $\Gamma^x_1=g^{x~2}_2\kappa/[\kappa^2+(\Delta^{x}_{c}-\delta_x)^2]$ and $\Gamma^y_1=g^{y~2}_2\kappa/[\kappa^2+(\Delta^{y}_{c}+\delta_x)^2]$. The value of $|\gamma|$ is larger when the
biexciton decay rates into both exciton states are equal, i.e., $\Gamma^x_1=\Gamma^y_1$. For positive $\delta_x=\omega_x-\omega_y$, if we choose $\Delta_c^x$ negative and $\Delta_c^y$ positive, the transition $|u\rangle\rightarrow|x\rangle$ and $|u\rangle\rightarrow|y\rangle$ will be detuned with cavity modes by $-(\Delta_c^x+\delta_x)$ and $\Delta_c^y+\delta_x$.
Because of the larger detunings, the decay rates of the biexciton states becomes smaller which enhances the entanglement between the generated photons. In addition, the ratio $\Gamma^x_1/\Gamma^x_2$ and $\Gamma^y_1/\Gamma^y_2$ is a maximum for cavity mode frequencies resonant with the excitons, i.e. $\Delta^{x}_{c}=\Delta^{y}_{c}=0$, and for larger values of $\delta_x$. In Figs.~\ref{fig3}(a,c), the cavity modes are resonant with the exciton frequencies and interact with the QD in
the weak coupling regime. The maximum possible values of $|\gamma|$ is nearly 0.35 (Fig.~\ref{fig3}(c)) which is close to the theoretical maximum  value of 0.367. For the strong coupling regime, the two frequencies of photons in each polarization become inseparable for small detunings. However, for larger detunings, when the photons are spectrally well resolved (see Figs.~\ref{fig3}(b,d)), the decay rates of the biexciton to excitons and the excitons to the ground state remains nearly
the same and the value of $|\gamma|$ is around 0.25.

\begin{figure}[t!]
\centerline{\includegraphics[width=3in]{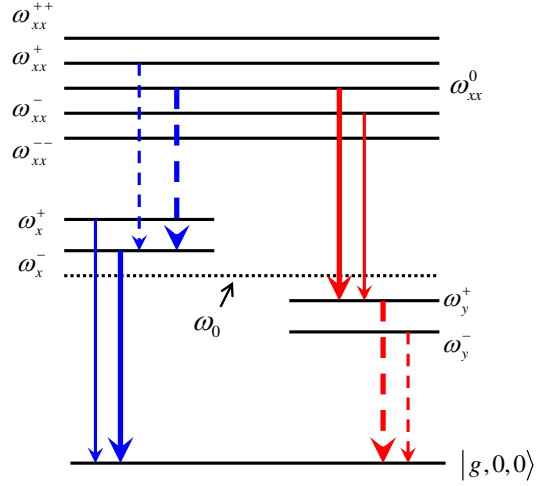}}
\caption{(color online)
The dressed states of biexciton and exciton for {\em across} generation of entangled photons. The upper five states $\omega_{xx}^{i}$ are the dressed state of the biexciton and the lower states $\omega_x^{i}$,  $\omega_y^i$ are the dressed state of the $|x\rangle$ and $|y\rangle$ excitons, respectively. The bold arrows  corresponding to the dominating peaks in the emitted spectrum.}
\label{eigen1}
\end{figure}

To  better understand the physical origin of the spectrum of Fig.~\ref{fig3}(b), we have analytically calculated the dressed states of the biexciton and excitons in the rotating frame with frequency $\omega_0$, in the strong coupling regime. We relegate the details
 of the calculation to the appendix. For an initial state $|u,0,0\rangle$, the coupled  cavity-QD system has five dressed states that can be expressed as the orthonormal superpositions of the bare states $|u,0,0\rangle$, $|x,1,0\rangle$, $|y,0,1\rangle$, $|g,2,0\rangle$, and $|g,0,2\rangle$. For $\Delta_c^x=-\Delta_c^y=\Delta$, $g_1^x=g_1^y=g_1$, and $g_2^x=g_2^y=g_2$, the energies of these biexciton dressed states are given by $\omega_{xx}^{0}=0$, $\omega_{xx}^{\pm}=\pm\sqrt{A-B}$, and $\omega_{xx}^{\pm\pm}=\pm\sqrt{A+B}$, where $A=[4g_2^2+(2\delta_x-3\Delta)^2+\Delta^2+8g_1^2]/4$, and $B=\sqrt{[2g_2^2+\Delta(2\delta_x-3\Delta)]^2+8g_1^2(2\delta_x-3\Delta)^2}/2$. After emitting the first photon via the leaky cavity mode, the system jumps to the dressed states of the excitons, which are superposition of either $|x,0,0\rangle$ and $|g,1,0\rangle$ or $|y,0,1\rangle$ and $|g,0,1\rangle$, depending on whether the emitted photon was $x-$polarized or $y-$polarized, respectively. The frequencies of the exciton dressed states are given by $\omega_x^{\pm}=(\delta_x-\Delta\pm\sqrt{4g_1^2+\Delta^{2}})/2$, $\omega_y^{\pm}=(-\delta_x+\Delta\pm\sqrt{4g_1^2+\Delta^{2}})/2$. In principle, the first emitted photon from the dressed states of biexciton can have ten peaks in the spectrum; however, for the initial state $|u,0,0\rangle$ and
 an off-resonant leaky cavity modes, only two peaks appear in the spectrum corresponding to $\omega_{xx}^0\rightarrow\omega_x^{-}$ and $\omega_{xx}^+\rightarrow\omega_{x}^-$ for $x-$polarized and $\omega_{xx}^0\rightarrow\omega_y^{+}$ and $\omega_{xx}^-\rightarrow\omega_{y}^+$
 for $y-$polarization; other possible transitions are negligible (see Fig.~\ref{eigen1}). Further, the peaks corresponding to $\omega_{xx}^0\rightarrow\omega_x^{-}$ and $\omega_{xx}^0\rightarrow\omega_y^{+}$ dominate completely. The second photon is emitted from the decay of the dressed states of excitons and have a two-peak spectrum corresponding to frequencies $\omega_{x}^{\pm}$ or $\omega_y^{\pm}$. The peaks corresponding to frequencies $\omega_x^-$ for the $x-$polarized photon and $\omega_y^+$ for  the $y-$polarized photon are largely dominating.

Although the value of $|\gamma|$ is limited by $2/e$ in {\em across generation} of photons through time delay, nevertheless, the entanglement can be distilled by using a frequency filter having two narrow spectral windows of width $w$ centered at the frequencies of degenerate peaks in the spectrum of $x-$polarized and $y-$polarized photons, say, $\omega_1$ and $\omega_2$. Subsequently, the response of the spectral filter can be written as a projection operator of the following form
\begin{eqnarray}
F(\omega_k,\omega_l)=\left\{\begin{array}{cc}1,&{\rm for}
~|\omega_k-\omega_1|<w,\\1,&{\rm for}
~|\omega_l-\omega_2|< w,~~~~~~~~~\\
0,&{\rm otherwise.}~~~~~~~~~~~~~~~~~\end{array}\right.
\label{rf}
\end{eqnarray}
After operating on the wave function of the emitted photons (Eq.~(\ref{state})), by the spectral function $F(\omega_k,\omega_l)$ and tracing over the energy states~\cite{akopian}, we get the reduced density matrix of the filtered photon pairs in the polarization basis. The normalized off-diagonal element of the density matrix for filtered photons $\gamma$
can be computed by
integrating over the projection operator of the filter~\cite{akopian}. One has
\begin{equation}
\gamma=\frac{\int\int c^{x*}_{kl}c^y_{lk}W_{opt}(\omega_k,\omega_l)F(\omega_k,\omega_l) d\omega_kd\omega_l}{
\int\int \left [ |c^x_{kl}|^2F(\omega_k,\omega_l)+
|c^y_{lk}|^2F(\omega_k,\omega_l) \right ]d\omega_kd\omega_l}.
%
%\int\int|c^x_{kl}|^2F(\omega_k,\omega_l)d\omega_kd\omega_l+\int\int|c^y_{lk}|^2F(\omega_k,\omega_l)d\omega_kd\omega_l},
\label{filter}
\end{equation}
We show in Figs.\,\ref{fig3}(c,d) (red curves) that large values of $|\gamma|$ can be achieved by using a
spectral filter.
The higher values of $|\gamma|$ are achieved because of the fact that the photons along the tails in the spectrum do not get time reordered properly using a {\em linear} time delay and thus reduce the entanglement. We find that the entanglement can be distilled by using a frequency filter with two spectral windows centered at the frequencies $\omega_x$ and $\omega_y$ for the weakly coupling case and $\omega_x^-$ and $\omega_y^+$ for the strong coupling case.
Again it should be noted that the conditional probabilities after filtering,  for generating entangled photon pairs, are very large (80\% for Fig.\,\ref{fig3}(c) and 50\% for Fig.\,\ref{fig3}(d)) because of the fact that photons are selected around the degenerate spectral peaks not along the degenerate tails as performed in earlier works~\cite{akopian}, where the conditional probabilities
are much less (e.g., less than 5\% conditional probabilities for 80\% concurrence values).

\section{``Within generation'' of entangled photons}
For {\em within generation} of entangled photons, the $x-$polarized and $y-$polarized photons should match in energy
within the same generations. In this case we consider the exciton states, which have a small
 but non-zero FSS, to interact with cavity modes in strong coupling regime so that the
%coupled
system forms degenerate polariton states~\cite{johne,ours}.
Here we extend previous works~\cite{johne,ours} by
considering that the biexciton state is also  coupled with the same cavity modes by reducing
 the binding energy; however,
the biexciton to exciton transition is more off-resonant so that further splitting in the polariton states due to biexciton couplings is negligible.

The state of the photon pair emitted via cavity modes can be rewritten as
\begin{equation}
|\psi\rangle=\sum_{k,l}[c^x_{kl}(\infty)|1_k\rangle_x|1_l\rangle_x+c^y_{kl}(\infty)|1_k\rangle_y|1_l\rangle_y] ,
\label{wt}
\end{equation}
The coefficients $c$ are given by the previously calculated Eqs.~(\ref{fcmn})-(\ref{fcpmn}). For state (\ref{wt}), the off-diagonal density matrix elements in the polarization basis is written as
\begin{equation}
\gamma=\frac{\int\int c^{x*}_{kl}(\infty)c^y_{kl}(\infty)\,d\omega_kd\omega_l}{\int\int |c^{x}_{kl}(\infty)|^2\,d\omega_kd\omega_l+|c^y_{lk}(\infty)|^2\,d\omega_kd\omega_l}.
\label{gm}
\end{equation}
We consider a positive detuning $\Delta_c^x$  and a negative detuning $\Delta_c^y$, which are equal to the FSS, i.e., $\Delta_c^x=-\Delta_c^y=\delta_x$. In this case, the biexciton to exciton transition $|u\rangle\rightarrow|x\rangle$ and $|u\rangle\rightarrow|y\rangle$ are equally detuned by $-\Delta_{xx}$. The exciton coupled with cavity modes form degenerate polariton states for $\Delta_c^x=-\Delta_c^y=\delta_x$. It should be noted here that although the biexciton is more detuned, still the decay rate of biexciton via cavity modes could be much larger than $\gamma_1$ (sum of radiative and nonradiative half width in free space).

\begin{figure}[t]
\centerline{\includegraphics[width=3.5in, height=2.5in]{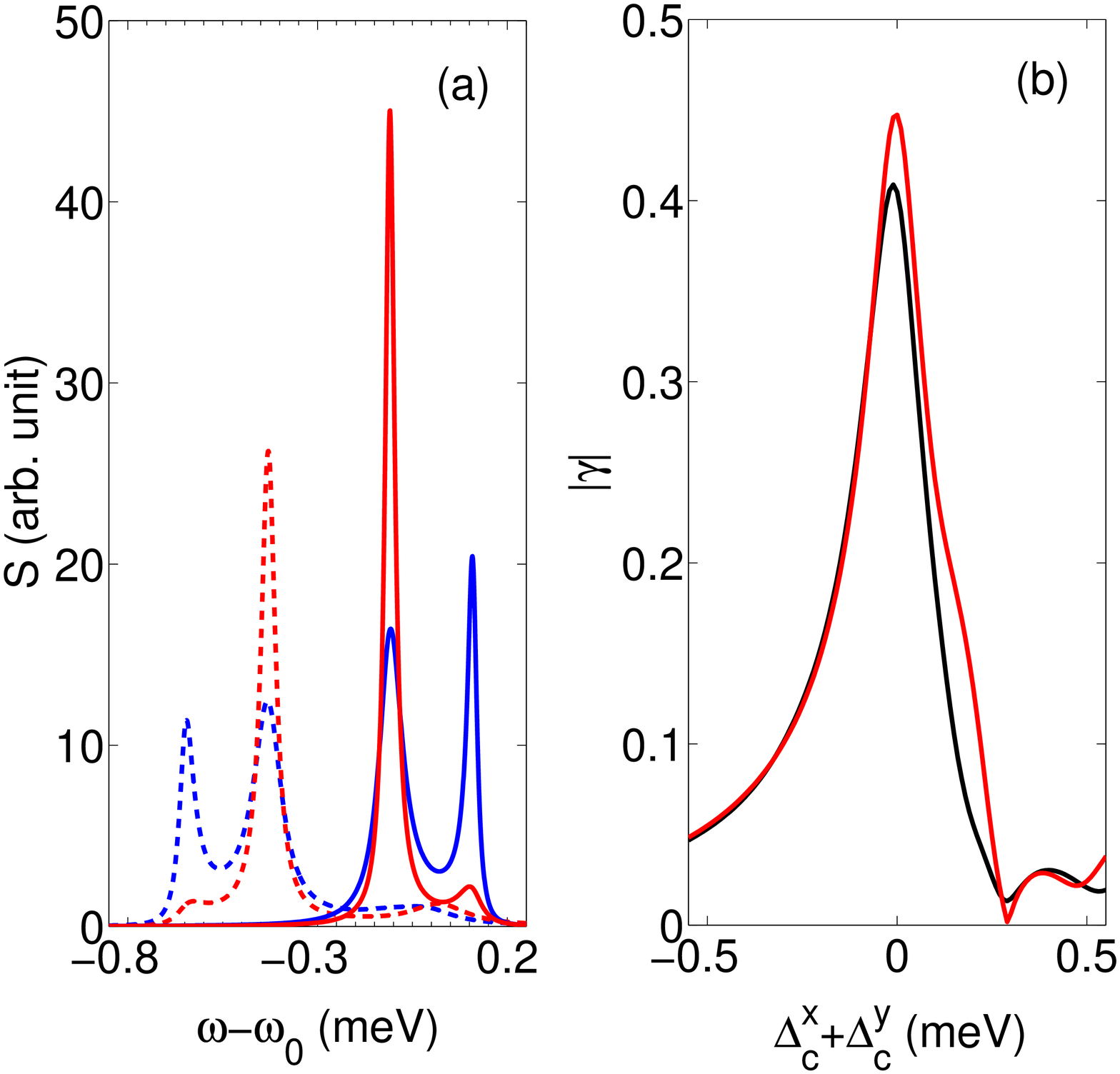}}
\vspace{-0.3cm}
\caption{(Color online)
The ``within generation'' of entangled photons when the biexciton is also coupled with the cavity modes; the
 biexciton binding energy is reduced to  $\Delta_{xx}=0.5$\,meV. (a) The spectrum of the photons $S(\omega)$ for $\delta_x=0.1$\,meV, $\gamma_1=2\gamma_2=0.004\,$meV, $\kappa=0.05$\,meV, $g^x_1=g^x_2=g^y_1=g^y_2=g=0.1$\,meV, and $\Delta_c^x=-\Delta_c^y=0.1$\,meV . The $x-$polarized photons are shown in blue and the $y-$polarized are shown in red; also, the solid curves are for photons generated in the exciton decay and the dotted curves are for photons generated in the biexciton decay.
(b) The values of $|\gamma|$ for generated photons, by changing $\Delta_c^x$ for $\Delta_c^y=-0.1$\,meV. The red (black) curve represents the results for filtered (unfiltered) photons; the filter function corresponds to two spectral windows of width $w=0.15\,$\,meV, centered at $\omega-\omega_0=-0.45$\,meV and $\omega-\omega_0=-0.05$\,meV.}
\vspace{-0.1cm}
\label{fig5}
\end{figure}
\begin{figure}[b!]
\centerline{\includegraphics[width=3in]{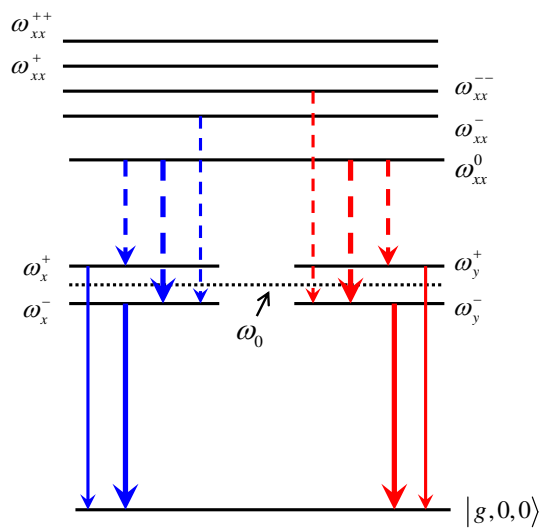}}
\caption{(color online)
Same as in Fig.~\ref{eigen1}, but for {\em within generation} of entangled photons.}
\label{eigen2}
\end{figure}

In Fig.~\ref{fig5}(a), we show the spectrum of the photons generated in first generation (dotted lines) and in the second generation (solid line). It is necessary that the first generation and the second generation photons should be well resolved spectrally, therefore a moderate ($\sim2\sqrt{4g_1^{x~2}+\delta_{x}^2}$) binding energy of the biexciton is essential for the within generation scheme of entangled photons. In this case, for $\Delta_c^x=-\Delta_c^y=\delta_x$, $g_1^x=g_1^y=g_1$, $g_2^x=g_2^y=g_2$, and $\Delta_{xx}\gg g_2$, we find that the dressed states of biexciton are given by (see appendix)
\begin{eqnarray}
\omega_{xx}^0&\approx&-\left(\Delta_{xx}+\frac{2g_2^2}{\Delta_{xx}}\right),  \\
\omega_{xx}^{+}&\approx&\epsilon^{+}+\frac{g_2^2\cos^2\theta}{\Delta_{xx}+\epsilon^{+}}, \\
\omega_{xx}^{-}&\approx&\epsilon^{-}+\frac{g_2^2\sin^2\theta}{\Delta_{xx}+\epsilon^{-}}, \\
\omega_{xx}^{++}&\approx&-\epsilon^{-}+\frac{g_2^2\sin^2\theta}{\Delta_{xx}-\epsilon^{-}}, \\
\omega_{xx}^{--}&\approx&-\epsilon^{+}+\frac{g_2^2\cos^2\theta}{\Delta_{xx}-\epsilon^{+}},
\end{eqnarray}
where $\epsilon_{\pm}=(-\delta_x\pm\sqrt{\delta_x^2+8g_1^2})/2$, and $\theta=\tan^{-1}[2\sqrt{2}g_1/(\delta_x+\sqrt{\delta_x^2+8g_1^2})]$. Using the parameters of Fig.~\ref{fig5}, $\omega_{xx}^0=-0.54$\,meV, $\omega_{xx}^+=0.11$\,meV, $\omega_{xx}^-=-0.19$\,meV, $\omega_{xx}^{++}=0.20$\,meV, $\omega_{xx}^{--}=-0.08\,$meV, and the dressed states of exciton are given by $\omega_x^{\pm}=\omega_y^{\pm}=\pm(4g_1^2+\delta_x^2)/2=\pm0.11$. Both exciton states for different polarization becomes degenerate for $\Delta_c^x=-\Delta_c^y=\delta_x$ \cite{johne,ours}. The schematic diagram of the dressed states is shown in Fig.\ref{eigen2}. The spectra of the first generation photons, mostly generated in the decay of biexciton dressed state $\omega_{xx}^0$, have two pronounced peaks corresponding to the frequencies $\omega_{xx}^0-\omega^{\pm}_{x}$, i.e., at -0.65 meV and -0.43 meV in Fig.~\ref{fig5}; there is a very small probability for generation of photons in the transitions $\omega_{xx}^-\rightarrow\omega_x^-$ and $\omega_{xx}^{--}\rightarrow\omega_y^{-}$ corresponding to frequencies -0.08 meV, 0.03 meV, respectively. The spectra of photons in the second generation have two peaks corresponding to dressed state of excitons at $\pm0.11$\,meV.

The calculated value of $|\gamma|$ is shown in Fig.~\ref{fig5}(b). For tuning the cavity mode frequencies, we fix one of the detunings $\Delta_c^x$ and $\Delta_c^y$, and change the other. This type of tuning has been
experimentally shown using AFM oxidization techniques~\cite{HennessyAPL06}, and note note
that this scheme would be suitable to
tune a large number of cavity-QD systems on the same chip. For this within generation
study, we find very large values of $|\gamma|$ for the {\em deterministic} generation of photons. For further distilling the entanglement, spectral filters can also be used, but
with a reduced probability and efficiency. Using spectral filtering, the maximally entangled photons can be generated with a small reduction of probability of detection. We show the results for spectrally filtered photons in Fig.~\ref{fig5}(b) by the red curve. The values of $|\gamma|$ are calculated using Eq.(\ref{gm}) after multiplying with the filter function (Eq.\,\ref{rf}).

\section{Conclusions}
In summary, we have presented methods for across generation and within generation of entangle photons using single QD coupled with a PC-cavity, and exploited the fact that the biexciton binding energy can be tuned. For zero biexciton binding energy, the concurrence for the across generation through time delay of photons is limited by 2/e, which can be enhanced to 1 using a spectral filter, at the expense of reduced probability. For small biexciton binding energies, the system can be tuned for efficient within generation of entangled photons. The concurrence larger than 0.8 has been shown for within generation of fast entangled photons, even without spectral filtering.

\appendix
\section{Dressed states of the biexciton}
The Hamiltonian for the system of the QD coupled with two-modes in PC-cavity, in the rotating frame with frequency $\omega_0=(\omega_x+\omega_y)/2$, for $\Delta_c^x=-\Delta_c^y=\Delta$, $g_1^x=g_1^y=g_1$, and $g_2^x=g_2^y=g_2$, and neglecting the coupling with environment, can be  written as
\begin{eqnarray}
\label{hr}
\frac{H_R}{\hbar} &  = & -\Delta_{xx}|u\rangle\langle u|
+\frac{\delta_x}{2}\left(|x\rangle\langle x|-|y\rangle\langle y|\right)
\nonumber \\
&&-\left(\Delta-\frac{\delta_x}{2}\right)\hat a_c^{x\dag}\hat a_c^x+\left(\Delta-\frac{\delta_x}{2}\right)\hat a_c^{y\dag}\hat a_c^y \nonumber \\
&&+[g_1|x\rangle\langle g|\hat a_c^x
+g_2|u\rangle\langle x|\hat a_c^x+g_1|y\rangle\nonumber\\
&\hbox{}&\langle g|\hat a_c^y+g_2|u\rangle\langle y|\hat a_c^y+H.c.].
\end{eqnarray}
For the across generation of entangled photons, $\Delta_{xx}=0$, we diagonalize the Hamiltonian and find the dressed energy states of the biexciton as follows
\begin{eqnarray}
&&\omega_{xx}^0=0,\\
&&\omega_{xx}^{+}=\sqrt{A-B},\\
&&\omega_{xx}^{-}=-\sqrt{A-B},\\
&&\omega_{xx}^{++}=\sqrt{A+B},\\
&&\omega_{xx}^{--}=-\sqrt{A+B},\\
\end{eqnarray}
with
\begin{eqnarray}
&&A=\frac{1}{4}[4g_2^2+(2\delta_x-3\Delta)^2+\Delta^2+8g_1^2]\nonumber\\
&&B=\frac{1}{2}\sqrt{[2g_2^2+\Delta(2\delta_x-3\Delta)]^2+8g_1^2(2\delta_x-3\Delta)^2}.\nonumber
\end{eqnarray}

For within generation of entangled photons, $\Delta_{xx}\neq0$ and $\Delta=\delta_x$, from Eq.(\ref{hr}) we can rewrite the Hamiltonian $H_R$,  in the basis of the state of the combined QD-cavity system as follows
\begin{eqnarray}
H_R&=&-\hbar\Delta_{xx}|u,0,0\rangle\langle u,0,0|+\hbar g_2\left[|u,0,0\rangle\langle x,1,0|\right.\nonumber\\&&\left.+|u,0,0\rangle\langle y,0,1|+H.c.\right]+H_S,\\
H_S&=&-\hbar\delta_x\left(|g,2,0\rangle\langle g,2,0|-|g,0,2\rangle\langle g,0,2|\right)\nonumber\\&&+\hbar g_1\sqrt{2}\left[|x,1,0\rangle\langle g,2,0|+|y,0,1\rangle\langle g,0,2| \right . \nonumber\\
& \hbox{}& \left . +H.c.\right].
\end{eqnarray}
After diagonalizing $H_S$, the eigenstates and corresponding eigenvalues of $H_S$ are given by
\begin{eqnarray}
|x_+\rangle & = & \cos\theta|x,1,0\rangle+\sin\theta|g,2,0\rangle,~~~~~\epsilon_+\\
|x_-\rangle & = & -\sin\theta|x,1,0\rangle+\cos\theta|g,2,0\rangle,~~\epsilon_-\\
|y_+\rangle & = & \sin\theta|y,0,1\rangle+\cos\theta|g,0,2\rangle,~~-\epsilon_-\\
|y_-\rangle & = & \cos\theta|y,0,1\rangle-\sin\theta|g,0,2\rangle,~~-\epsilon_+,
\end{eqnarray}
where $\epsilon_{\pm}=(-\delta_x\pm\sqrt{\delta_x^2+8g_1^2})/2$, and $\theta=\tan^{-1}[2\sqrt{2}g_1/(\delta_x+\sqrt{\delta_x^2+8g_1^2})]$.
We can rewrite the Hamiltonian $H_0$ in terms of eigenstates of $H_s$ as follows
\begin{eqnarray}
H_0&=&-\hbar\Delta_{xx}|u\rangle\langle u|+\hbar\epsilon^+(|x_+\rangle\langle x_+|-|y_-\rangle\langle y_-|)\nonumber\\&&+\hbar\epsilon^-(|x_-\rangle\langle x_-|-|y_+\rangle\langle y_+|)\nonumber\\
&&+\hbar g_2\cos\theta\left[|u\rangle\langle x_+|+|u\rangle\langle y_-|+H.c\right]\nonumber\\
&&-\hbar g_2\sin\theta\left[|u\rangle\langle x_-|-|u\rangle\langle y_+|+H.c\right].
\end{eqnarray}
For $\Delta_{xx}>>g_2$, we can use perturbation theory and obtain the frequency eigenvalues
\begin{eqnarray}
\omega_{xx}^0&=&-\Delta_{xx}-2\Delta_{xx}\left(\frac{g_2^2\cos^2\theta}{\Delta_{xx}^2-\epsilon^{+2}}
+\frac{g_2^2\sin^2\theta}{\Delta_{xx}^2-\epsilon^{-2}}\right) , \nonumber\\
&\approx&-\left(\Delta_{xx}+\frac{2g_2^2}{\Delta_{xx}}\right) , \\
\omega_{xx}^{+}&=&\epsilon^{+}+\frac{g_2^2\cos^2\theta}{\Delta_{xx}+\epsilon^{+}} , \\
\omega_{xx}^{-}&=&\epsilon^{-}+\frac{g_2^2\sin^2\theta}{\Delta_{xx}+\epsilon^{-}} ,\\
\omega_{xx}^{++}&=&-\epsilon^{-}+\frac{g_2^2\sin^2\theta}{\Delta_{xx}-\epsilon^{-}} , \\
\omega_{xx}^{--}&=&-\epsilon^{+}+\frac{g_2^2\cos^2\theta}{\Delta_{xx}-\epsilon^{+}}.
\end{eqnarray}

\end{document}